\iftrue
\documentclass[aps,prx,twocolumn,
			   groupedaddress,superscriptaddress,
			10pt,   amsfonts,amssymb,amsmath,
			   citeautoscript,
			   a4paper]{revtex4-2}
\else
\documentclass[aps,pra,preprint,
			   groupedaddress,superscriptaddress,
			   amsfonts,amssymb,amsmath,
			   citeautoscript,
			   a4paper]{revtex4-2}
\fi

\usepackage[pdftex]{hyperref}
\hypersetup{colorlinks,
            linkcolor={blue!75!black!80!yellow},
            citecolor={blue!75!black!80!yellow},
            urlcolor={blue!75!black!80!yellow},
            pdfstartview=FitH}
\usepackage{graphicx}
\usepackage{mathrsfs}
\usepackage{xspace}
\usepackage{braket}
\usepackage{xr}
\usepackage{gensymb} 
\usepackage{xcite}
\usepackage{xcolor,soul}
\usepackage{stmaryrd} 
\usepackage[UKenglish]{babel}
\usepackage{placeins} 
\usepackage{physics}

\usepackage{siunitx}
\DeclareSIUnit[number-unit-product=]\percent{\char`\%} 

\usepackage{txfonts}  
\usepackage{txfontsb} 

\makeatletter
\newcommand*{\addFileDependency}[1]{
  \typeout{(#1)}
  \@addtofilelist{#1}
  \IfFileExists{#1}{}{\typeout{No file #1.}}
}
\makeatother

\newcommand{\appropto}{\mathrel{\vcenter{
			\offinterlineskip\halign{\hfil$##$\cr
				\propto\cr\noalign{\kern2pt}\sim\cr\noalign{\kern-2pt}}}}}

\usepackage{textcomp} 
\usepackage{xifthen}
\usepackage{etoolbox}
\newboolean{togglecomments}
\newboolean{togglechanges}
 
\setboolean{togglecomments}{true}
\setboolean{togglechanges}{false}

\newcommand{\comment}[2]{%
    \ifbool{togglecomments}%
    {\textcolor{blue!70!black}{\small\textsf{%
    \textsuperscript{\textsc{\textsf{\MakeLowercase{#1}}}}%
    [#2]}}} 
    {}}     
\newcommand{\swap}[2]{\ifbool{togglechanges}
    {#2}  
    {\textcolor{red!70!black}{[#1]}\textrightarrow{}\textcolor{green!50!black}{[#2]}}}
\newcommand{\remove}[1]{\ifbool{togglechanges}
    {}    
    {\textcolor{red!70!black}{#1}}}
\newcommand{\inset}[1]{\ifbool{togglechanges}
    {#1}  
    {\textcolor{green!50!black}{#1}}}
\newcommand{\optional}[1]{\ifbool{togglechanges}
    {}    
    {\textcolor{yellow!50!orange!80!gray}{#1}}}

\newcommand{\citeremind}[1]{%
    [\textcolor{blue!75!black!80!yellow}{
        $\blacksquare$%
	    \ifthenelse{\isempty{#1}}
	        {}
	        {\textsuperscript{\tiny\textsf{#1}}}%
	}]\xspace}

\newcommand{\hkuaffil}{\footnotesize Department of Physics and HK Institute of Quantum Science and Technology,\\The University of Hong Kong, Pokfulam, Hong Kong, China}

\begin{document}
\title{Non-Abelian Electric Field and Zitterbewegung on a Photonic Frequency Chain}

\author{Shu~Yang}
\affiliation{\hkuaffil}
\author{Bengy~Tsz~Tsun~Wong}
\affiliation{\hkuaffil}
\author{Yi~Yang}
\email{yiyg@hku.hk}
\affiliation{\hkuaffil}
\begin{abstract}
The synthetic frequency dimension, which realizes fictitious spatial dimensions from the spectral degree of freedom, has emerged as a promising platform for engineering artificial gauge fields in studying quantum simulations and topological physics with photons. 
A current central task for frequency-domain photons is the creation and manipulation of nontrivial non-Abelian field strength tensors and observing their governing dynamics.
Here, we experimentally demonstrate a miniaturized scheme for creating non-Abelian electric fields in a photonic frequency chain using a polarization-multiplexed, time-modulated ring resonator. 
By engineering spin-orbit coupling via modulation dephasing, polarization rotation, and polarization retardation, we achieve programmable control over synthetic Floquet bands and their quasimomentum spin-resolved textures. 
Leveraging self-heterodyne coherent detection, we demonstrate Zitterbewegung---a trembling motion of photons---induced by non-Abelian electric fields on the frequency chain. We further observe the interference between Zitterbewegung and Bloch oscillations arising from the coexistence of non-Abelian and Abelian electric fields. 
Our work bridges synthetic dimensions with non-Abelian gauge theory for versatile photonic emulation of relativistic quantum mechanics and spinor dynamics, and can be instrumental in applications like frequency-domain optical computation and multimodal frequency comb control.
\end{abstract}
\maketitle
Synthetic non-Abelian gauge fields are fundamental building blocks for exploring topological physics and quantum dynamics in engineered physical platforms~\cite{dalibard2011colloquium,goldman2014light,aidelsburger2018artificial,ozawa2019topological,banuls2020simulating,aidelsburger2022cold,yang2024non}.
One crucial distinction between Abelian and non-Abelian gauge fields lies in their field strength. 
In Abelian electrodynamics governed by the U(1) gauge group, the field strengths solely depend on the spatial and temporal derivatives of the gauge fields, which preclude self-interaction~\cite{jackson2012classical}.
Non-Abelian gauge theory generalizes this framework: electric fields and magnetic fields additionally depend on the non-commutativity between the matrix-valued scalar and vector potentials, enabling topological phenomena such as synthetic spin-orbit coupling (SOC)~\cite{ruseckas2005non,stanescu2007nonequilibrium,lin2011spin,zhai2015degenerate,zhang2015spin,huang2016experimental,wu2016realization,rechcinska2019engineering,sala2015spin,liang2024polariton,li2022manipulating,spencer2021spin,shi2025coherent}, non-trivial momentum spin texture~\cite{whittaker2021optical,terccas2014non,gianfrate2020measurement}, Zitterbewegung (ZB) dynamics~\cite{merkl2008atomic,vaishnav2008observing,leblanc2013direct,dreisow2010classical,qu2013observation,zhang2008observing,sedov2018zitterbewegung,hasan2022wave,zawadzki2011zitterbewegung,polimeno2021experimental,lovett2023observation,chen2019non}, non-Abelian Hofstadter physics~\cite{osterloh2005cold,goldman2009non,yang2020non}, and non-Abelian Aharonov-Bohm interference~\cite{wu1975concept,wilczek1984appearance,horvathy1986non,bohm2013geometric,yang2019synthesis,iadecola2016non,noh2020braiding,jacob2007cold,dong2024temporal,fruchart2020dualities,wu2022non}.

Photonic systems have recently emerged as an intriguing testbed for synthesizing non-Abelian fields~\cite{yang2024non,yan2023non}. 
Among these, the synthetic frequency dimension~\cite{yuan2018synthetic, ozawa2016synthetic, bell2017spectral, ozawa2019nrp, dutt2019experimental, dutt2020single, lustig2021topological, englebert2023bloch, wang2021generating, wang2021topological, chen2021real, villa2024mean, sridhar2025measuring, yu2025comprehensive}
offers exceptional scalability, programmability, and compatibility with free-space, fiber, and integrated photonic architectures. Equipped with precise linear and nonlinear control, it holds promises for on-chip applications such as spectral shaping of light, frequency comb manipulation, and frequency-domain optical computation ~\cite{hu2020realization,zhao2022enabling,balvcytis2022synthetic,hu2022mirror,dinh2024reconfigurable,wang2025versatile,zeng2025hybrid}.
Polarization-multiplexed time-modulated ring resonators are theoretically revealed to host frequency-domain non-Abelian fields across the Hermitian and non-Hermitian regimes~\cite{cheng2023artificial, pang2024synthetic,wong2025synthetic}
with a recent experimental demonstration of non-Abelian lattice gauge fields in a quasi-2D system under the twisted boundary condition~\cite{cheng2025non}.
However, existing research has primarily focused on non-Abelian magnetic fields for photons; the synthesis of their electric counterparts becomes crucial for the complete access and manipulation of the non-Abelian field strength tensor. Furthermore, fundamental phenomena like ZB dynamics induced by non-Abelian fields remain experimentally elusive in the photonic frequency dimension.

In this work, we experimentally realize a compact and versatile scheme for synthesizing non-Abelian electric fields on the photonic frequency chain based on a polarization-multiplexed time-modulated ring resonator.
We demonstrate the comprehensive control of the synthetic SOC via modulation dephasing, polarization rotation, and polarization retardation. 
Leveraging self-heterodyne coherent detection, we further unveil ZB dynamics and their intricate interplay with Bloch oscillations (BO), arising from the simultaneous presence of Abelian and non-Abelian electric fields. 
\begin{figure*}[htbp]
    \centering
    \includegraphics[width=1\linewidth]{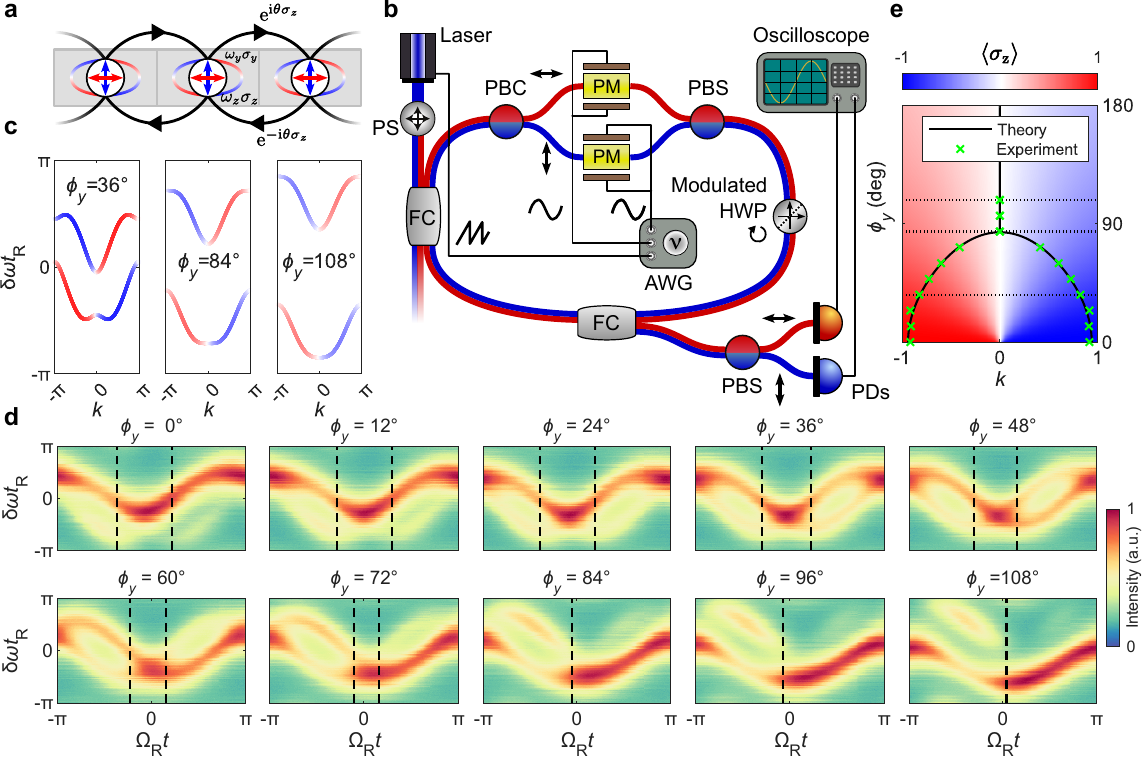}
    \caption{
    \textbf{Non-Abelian electric fields in the photonic frequency dimension. } 
    \textbf{a}.~One-dimensional lattice model featuring non-commutative matrix-valued scalar and vector potentials.
    \textbf{b}.~Simplified schematic of the experimental setup (see full setup in SM Sec.~S1A). The red and blue colors denote the horizontal and vertical polarization in the polarization-maintaining fibers (PMF), which are also labeled by the orientation of the double-sided arrows. 
    Black lines correspond to electronic cables. PS, polarization synthesizer; FC, fiber coupler; PBS/C, polarization beam splitter/combiner; HWP, half-wave plate; PM, phase modulator; AWG, arbitrary-wave generator; PD, photo-detector. 
    \textbf{c}.~Band structure and spin texture of the theoretical model. The bands feature a double-minima to single-minima transition as the rotation $\phi_y$ increases.
    \textbf{d}.~Total intensity from both polarizations showing band structure evolution under $\phi_y$ sweeping. Dashed lines indicate the quasi-momenta of the single-particle ground state. 
    \textbf{e}. Quasimomenta of the single-particle ground state overlaid with the corresponding spin-texture hotmap. The black solid line is the theoretical fitting for the experimental values (green crosses) obtained from d. The three horizontal dotted cuts correspond to the band structures in c. Here $\omega_z=0$, $\theta=0.31\pi$, $Jt_\mathrm{R}=-0.20\pi$. }
    \label{fig:fig1}
\end{figure*}

The non-Abelian electric field is the $01$-th component of the rank-two Yang-Mills field strength tensor:
\begin{align}
E^i = -\partial A^i /\partial t - \nabla_i V -\mathrm{i}[V, A^i]\,, \label{eq:nonAB_E_general}
\end{align}
where all gauge field variables $V$ and $A^{i}$ are matrices and we assume a $\mathrm{diag(-\!+\!++)}$ signature of the Minkowski metric. 
The time and spatial derivatives in Eq.~\eqref{eq:nonAB_E_general} are inherited from the definition of the Abelian electric field, whereas the last term in Eq.~\eqref{eq:nonAB_E_general} requires the non-commutativity between the scalar and vector potentials such that the non-Abelian part of the electric field can appear. 
To this end, we consider the lattice Hamiltonian (Fig.~\ref{fig:fig1}a)
\begin{align} \label{eq:Hamiltonian}
\begin{aligned}
\hat{H} = \sum_{m}\bigg[&-\omega_y \hat{a}^\dagger_{m} \sigma_y \hat{a}_{m}-\omega_z \hat{a}^\dagger_{m} \sigma_z \hat{a}_{m}\\
&+J\hat{a}_{m}^\dagger \mathrm{e}^{-\mathrm{i}\theta\sigma_z} \hat{a}_{m+1} + J\hat{a}^\dagger_{m+1}\mathrm{e}^{\mathrm{i}\theta\sigma_z}\hat{a}_{m}   \bigg]\,,
\end{aligned}
\end{align}
where $m$ indexes the lattice site, $\hat{a}^\dagger_{m}$ and $\hat{a}_{m}$ are the creation and annihilation
operators, and $\sigma_{x,y,z}$ are Pauli matrices.
%
Its corresponding Bloch Hamiltonian is 
\begin{equation} \label{eq:kHamiltonian}
H(k) = 2J\cos(k+\theta\sigma_z) -\omega_y \sigma_y-\omega_z \sigma_z \,,
\end{equation}
which faithfully realizes a minimally coupled synthetic lattice under an equal mixing of Rashba and Dresselhaus SOC (see SM Sec.~S7A). 
The first two on-site terms in Eq.~\eqref{eq:Hamiltonian} contribute to a non-Abelian on-site scalar potential defined as $V=(-\omega_y\sigma_y-\omega_z\sigma_z)$, while the remaining hopping terms contain a non-Abelian 
vector potential $A_x=(-\theta\sigma_z)$.
In this one-dimensional configuration, non-Abelian magnetic fields are absent for lack of flux; nevertheless, non-Abelian electric fields can still be defined as $\hat{E}_x =2\omega_y\theta\sigma_x$ according to Eq.~\eqref{eq:nonAB_E_general} due to the non-commutativity between the scalar and vector potentials, despite both being spatially homogeneous and temporally stationary. 

To synthesize this Hamiltonian, we realize a polarization-multiplexed ring-resonator system as shown in Fig.~\ref{fig:fig1}b, where we take the two orthogonal polarizations of photons as the pseudospins.
Without loss of generality, we have assigned the horizontal and vertical polarization as the pseudospin-up and pseudospin-down states, respectively.
The main body of the ring resonator carries three major components:
(i) A polarization-dependent co-sinusoidal phase modulation with a dephasing of $\pm\theta$ between the two polarization axes. 
It realizes the Peierls-substituted hopping in Eq.~\eqref{eq:kHamiltonian} containing the SU(2) vector potential with the transfer function given by $\mathrm{e}^{\mathrm{i}g\cos(\Omega t +\theta\sigma_z)}$, where the modulation frequency $\Omega$ is chosen near the ring's free spectral range (FSR) $\Omega_{\mathrm{R}}$ such that a one-dimensional spinful chain can be formed by the polarization-multiplexed, longitudinal modes of the ring.
The modulation depth $g$ relates to the hopping strength $J$ by $J\approx-g/2t_\mathrm{R}$ in the weak modulation limit $g\rightarrow0$. 
In the experiment, it is conducted by two parallel lithium-niobate--based phase modulators between a pair of polarization beam splitter/combiner.
(ii) A polarization rotation described by $\mathrm{e}^{\mathrm{i}\phi_y\sigma_y}$ where $\phi_y=\omega_y t_\mathrm{R}$ enables the coupling between the two polarizations, with $t_\mathrm{R}=2\pi/\Omega_\mathrm{R}$ being the round-trip time. 
This achieves the second term in Eq.~\eqref{eq:kHamiltonian} and is physically achieved by a tunable in-line half-wave plate (HWP) set at an angle of $\phi_y/2$.
(iii) By applying different DC biases to the phase modulators, a phase retardation of $\mathrm{e}^{\mathrm{i}\phi_z\sigma_z}$ where $\phi_z=\omega_z t_\mathrm{R}$ can be introduced between the two polarizations to realize the third term in Eq.~\eqref{eq:kHamiltonian}. 
(ii) and (iii) jointly realize the SU(2) scalar potential.
Outside the ring, the arbitrary input state is prepared via a tunable continuous-wave (CW) laser followed by a polarization synthesizer, and then injected into the ring resonator via a weakly coupled fiber coupler. 
The scanning across the synthetic dimension is achieved by applying a sawtooth signal to the tunable laser. 
The synthesizer, when placed outside the ring, can deterministically prepare arbitrary polarization as the input state, while inside the ring, it can function as a programmable HWP to modulate the photon polarization rotation at a maximum rate of \SI{100}{kHz}, which is crucial for our demonstration of the ZB dynamics and its $\phi_y$ dependence below.
The signals inside the ring are sampled through another coupler and separated by a polarization beam splitter, after which the intensities of each polarization can be obtained by a pair of photodetectors.
The whole experiment is realized in an inline polarization-maintaining manner, without utilizing single-mode fibers or free-space elements, to minimize uncontrollable polarization crosstalk and insertion loss. 
This approach eliminates the need for amplifiers inside the ring, thereby helping reduce the noise and enhance the coherence of the measurements. 

The SOC in the Hamiltonian leads to band gap opening and spin exchanges within the two bands as shown in Fig.~\ref{fig:fig1}c. 
In the absence of $\phi_z$, the lower band exhibits two energy-degenerate minima at $k=k_\pm$ for small $\phi_y$. 
As $\phi_y$ increases, the two minima eventually merge at $k=0$. 
Meanwhile, the spin texture also inverts upon traversing the avoided crossing of the two bands.
We directly measure such band evolution in Fig.~\ref{fig:fig1}d-e by probing the intensity of horizontally and vertically polarized light from the dropping coupler. 
In this measurement, we do not require a specified input state for the optimal projection onto the eigenstates. 
Therefore, we put the synthesizer inside the ring serving as an electronically modulated HWP to tune the polarization rotation $\phi_y$ at a repetition rate of \SI{100}{Hz} for a $0\degree$ to $108\degree$ scanning by a step of $12\degree$. 
The input laser frequency $\omega$ is detuned with a repetition rate of \SI{1}{kHz}, so that a full FSR (and hence a full band) can be scanned at every $\phi_y$. 
Fig.~\ref{fig:fig1}d displays all ten frames of the measurements.
For each measurement, we trace the energy bands and locate their ground-state quasimomenta $k\equiv\Omega_\mathrm{R}t$ (see details in SM Sec.~S2), which are marked by one or two vertical dashed line(s) for each panel in Fig.~\ref{fig:fig1}d. 
Such a collection of measured $k_\pm$ are plotted in Fig.~\ref{fig:fig1}e as green crosses, which can be well described by the theoretical expression of $k_\pm =\pm \sin^{-1}\sqrt{\sin^2\theta - \omega_y^2\cot^2\theta/4J^2}$~\cite{wong2025synthetic}, where the parameters are fitted as $\theta=0.3\pi$ and $Jt_\mathrm{R}=-0.2\pi$ (solid line in Fig.~\ref{fig:fig1}e).
The fitted $\theta$ and $J$ exhibit small offsets from their nominal values of $\theta=0.25\pi$ and $Jt_\mathrm{R}=-0.25\pi$, which likely results from the breakdown of the first-order approximation (see SM Sec.~S1C). 

\begin{figure*}[htbp]
    \centering
    \includegraphics[width=1\linewidth]{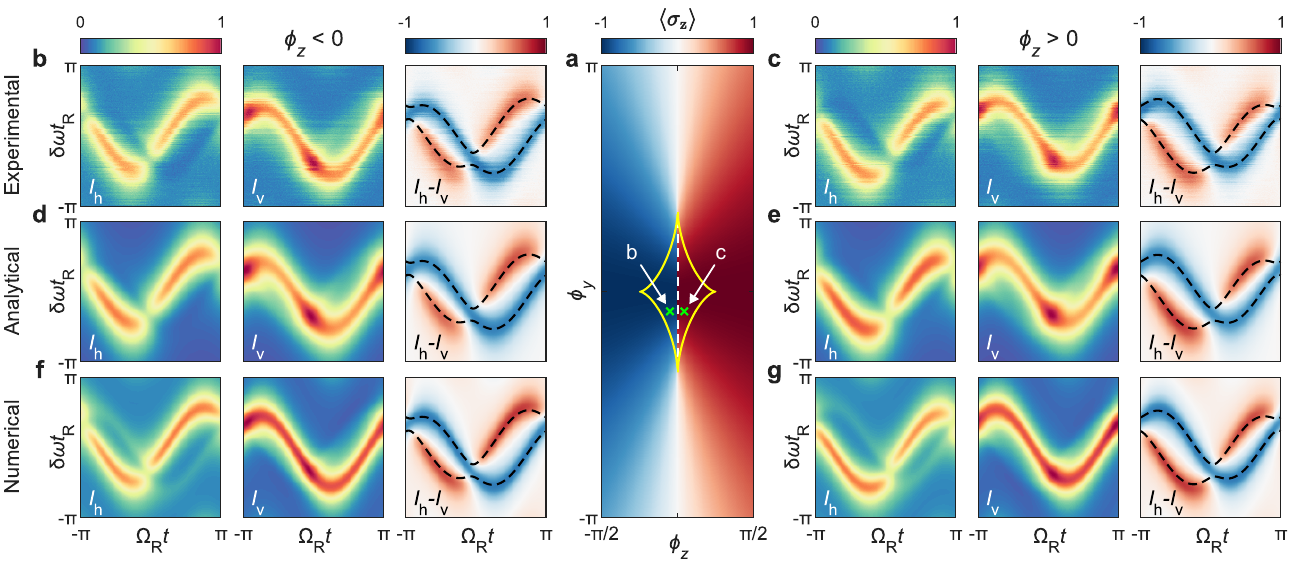}
    \caption{
    \textbf{Spin textures under synthetic spin-orbit coupling in the Floquet band structure.}
    \textbf{a.}~Theoretical single-particle ground-state phase diagram overlaid with spin textures. Outside and inside the yellow solid curve, the SOC band possesses single and double local minima, while the white dashed line marks the ground-state degeneracy. The green crosses correspond to the measurements in b and c.
    \textbf{b-c.}~Experimentally measured, \textbf{d-e.}~analytically calculated, and \textbf{f-g.}~numerically simulated intensity spectra of the horizontal polarization (left), vertical polarization (middle), and their differences (right).
    The dashed curves depict the theoretical bands. 
    A negative $\phi_z$ leads to a positive ground-state quasi-momentum in b, d, and f, whereas a positive $\phi_z$ leads to a negative ground-state quasi-momentum in c, e, and g. 
    The experimental parameters, directly applied to the numerical simulations, are $\theta=0.25\pi$, $\phi_y=-0.09\pi$, $g=0.5\pi$, and $\gamma_\mathrm{R}t_\mathrm{R}=0.63$, where the latter two are slightly modified as $g=0.47\pi$ and $\gamma_\mathrm{R}t_\mathrm{R}=0.83$ for the analytical calculation due to the violation of first-order approximations.
    $\phi_z=-0.05\pi$ for b, d and f, and $\phi_z=0.04\pi$ for c, e and g. 
    }
    \label{fig:fig2}
\end{figure*}

We demonstrate the spin texture exchanges of the two bands and the control of the single-particle ground state in Fig.~\ref{fig:fig2}. 
We adopted the spectroscopic scanning of the laser detuning $\delta\omega$ again, but this time with $\phi_z$ turned on. 
Because the rotation $\phi_y$ is fixed during this measurement, we replaced the synthesizer with a mechanical in-line HWP and moved the synthesizer outside the ring to prepare an input state that balances the projection of both polarizations. 
The retardation $\phi_z$ contains the contributions from both the birefringence of the polarization-maintaining fibers and the DC biases applied to the phase modulators.
The fiber's intrinsic birefringence can be large, but only its value modulo $2\pi$ is relevant and can be treated as a constant during the measurement window since it is slow-varying. 
Hence, the total $\phi_z$ can be well compensated and controlled by the phase modulators. 
In the experiment, two non-zero $\phi_z$ of opposite signs are chosen on the phase diagram (Fig.~\ref{fig:fig2}a), where the corresponding degeneracy-lifted new ground state is predicted to occur at positive and negative quasimomenta, respectively. 
Fig.~\ref{fig:fig2}b shows the time-resolved intensity of the two polarizations and their contrast defined by $I_\mathrm{h}-I_\mathrm{v}$ for $\phi_z<0$. 
Evidently, the two bands exchange their spin polarization, and the degeneracy of the ground state is lifted, with the ground-state quasimomentum being positive.
Conversely, when a positive $\phi_z$ is chosen, the ground-state quasimomentum becomes negative, as observed in Fig.~\ref{fig:fig2}c. 
In both Fig.~\ref{fig:fig2}b and Fig.~\ref{fig:fig2}c, $I_\mathrm{h}$ exhibits a distinct linewidth and intensity reduction near the avoided crossing, and conversely for $I_\mathrm{v}$. 
These features, absent in the low-loss limit, are consequences of loss-induced band broadening and can be modified upon the choice of the input state (see details in SM Sec.~S3).

We perform analytical (Fig.~\ref{fig:fig2}d-e) and numerical (Fig.~\ref{fig:fig2}f-g) validations to the measurements (Fig.~\ref{fig:fig2}b-c). 
Under the first-order approximation, the analytical intensity projections are given by 
$
I_\alpha\propto\bigg| \sum_{j=1,2}  { \braket{\psi_\alpha | \psi_j } \braket{\psi_j |\psi_\mathrm{in}} \bigg/ \left( E_j -\delta\omega - \mathrm{i}\gamma_\mathrm{R}\right)} \bigg| ^2 \,,
$
where $\alpha=\left\{\mathrm{h},\mathrm{v}\right\}$, $\ket{\psi_{1,2}}$ are the eigenvectors of the Hamiltonian in Eq.~\eqref{eq:kHamiltonian}, $\ket{\psi_\mathrm{in}}$ is the input polarization state, $\delta\omega$ is the frequency detuning of the laser, $\gamma_\mathrm{R}$ is the round-trip loss rate.
$E_{1,2}$ are the $k$-dependent eigen-energies given by $E_{\pm}(k) = 2J\cos\theta\cos k \pm \sqrt{\omega_y^2 + \omega_z^2 +4J^2 \sin^2\theta\sin^2 k +4J\omega_z \sin\theta\sin k }$. 
On the other hand, we develop a full-wave simulation tool that can comprehensively analyze the system response beyond the first-order approximation (see SM Sec.~S8), which has not been realized in previous studies of the photonic synthetic frequency dimension. 
The method faithfully simulates the time-domain response via the transfer function $\mathbf{E}(t+t_{\mathrm{R}})=\mathbf{T}\cdot\mathbf{E}(t)$, where $\mathbf{E}=(E_\mathrm{h}, E_\mathrm{v})^\mathrm{T}$ is the photonic pseudospin and the transfer function $
\mathbf{T} = \mathrm{e}^{\mathrm{i}\omega t_\mathrm{R}-\gamma_\mathrm{R}t_{\mathrm{R}}}\mathrm{e}^{\mathrm{i}g \cos(\Omega t+\theta\sigma_z)}\,\mathrm{e}^{\mathrm{i}\phi_y\sigma_y}\mathrm{e}^{\mathrm{i}\phi_z\sigma_z}$.
Excellent agreement is achieved among the experimental data, the analytical calculations, and the numerical simulations in Fig.~\ref{fig:fig2}.

\begin{figure*}[htbp]
    \centering
    \includegraphics[width=1\linewidth]{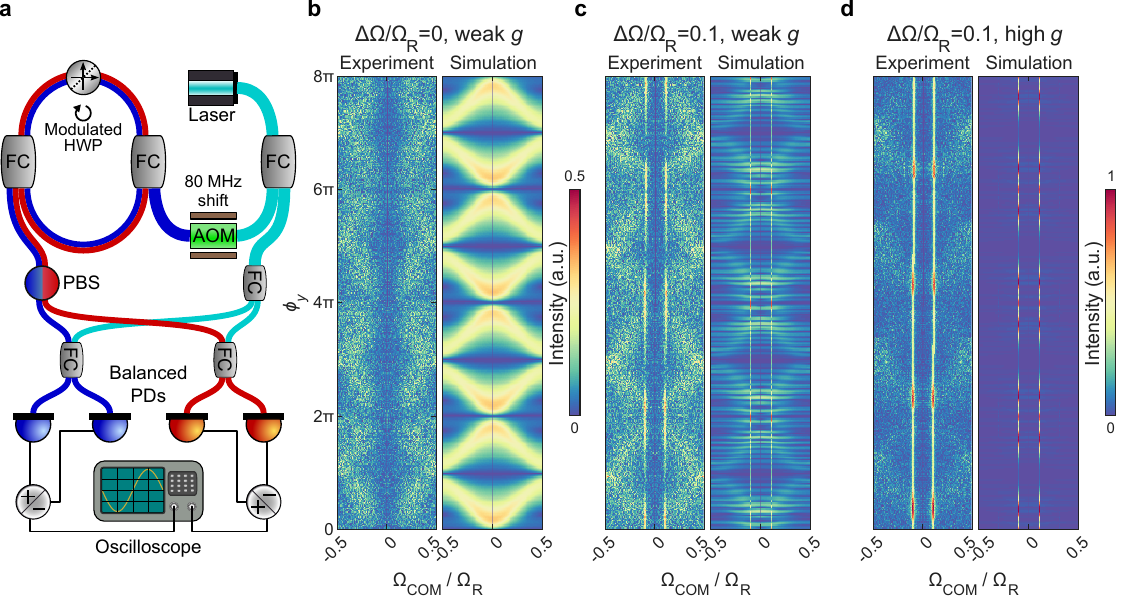}
    \caption{\textbf{Self-heterodyne coherent measurement of Zitterbewegung and Bloch oscillation in spectral dynamics.}
    \textbf{a}.~Simplified schematic of the coherent detection setup.
    The rotation $\phi_y$ modulation is realized with the polarization synthesizer placed inside the ring, serving as a programmable HWP. 
    Other optical elements inside the ring are omitted (see full setup in SM Sec.~S1B). 
    The signals were shifted by \SI{80}{MHz} using an acousto-optic modulator before entering the ring. The output lights are sent to a pair of balanced PDs and interfere with the unshifted self-heterodyne reference light. 
    \textbf{b.} Pure ZB effect under a weak modulation depth and zero modulation detuning.
    \textbf{c.} Simultaneous presence of ZB and BO and their interference under detuned weak modulation.
    \textbf{d.} BO dominance under detuned strong modulation. 
    All measurements are performed under input state $(0,1)^{\mathrm{T}}$, retardation $\phi_z$ is calibrated near zero, and modulation dephasing $\theta=0.25\pi$. 
    In all spectra, the irrelevant strong DC line at $\Omega_\mathrm{COM}=0$ is removed for better data presentation.}
    \label{fig:fig3}
\end{figure*}
A profound consequence of the synthesized non-Abelian electric fields is the emergence of ZB dynamics, which manifests as an oscillatory motion of the center of mass (COM) of the spectral distribution of the optical field amplitude. 
Crucially, ZB differs from BO---another oscillatory dynamics but induced by Abelian electric fields~\cite{yuan2016bloch}---in its modulation detuning dependence. 
While BO necessitates a non-zero frequency detuning to occur, ZB persists regardless of detuning, highlighting a fundamental distinction between the two dynamics in the frequency dimension. 

We are able to measure the ZB and BO phenomena and their interference in the spectral dynamics.
Since this oscillation appears in the amplitude's frequency distribution, simple intensity detection in Fig.~\ref{fig:fig1}b is no longer sufficient. 
To directly resolve the electric field, we upgraded the detection scheme to a self-heterodyne coherent detection using balanced PD pairs. 
Additionally, the balanced PDs also eliminate the common-mode noise, hence improving the signal-to-noise ratio such that we can measure the weak ZB oscillations.
As shown in Fig.~\ref{fig:fig3}a, 1\% of the source laser power is split off to serve as the self-heterodyne reference, with the remainder shifted by \SI{80}{MHz} using an acousto-optic modulator before entering the ring.
The two polarization components from the ring output are separated by a polarization beam splitter, and then each of them coherently interacts with the self-heterodyne reference in the 50:50 coupler and finally gets detected by a balanced photo-detector pair (details in SM Sec.~S4). 

Because the non-Abelian electric field is a function of $\omega_y$, we can rapidly modulate $\phi_y=\omega_yt_\mathrm{R}$ using the programmable synthesizer placed inside the ring, and monitor the change of the center of mass $\Omega_{\mathrm{COM}}$ induced by ZB along the frequency chain (see SM Sec.~S5).
During the experiment, we swept the polarization rotation $\phi_y$ from $0$ to $2\pi$ over four full cycles. 
We first measure the pure ZB phenomena in the absence of modulation frequency detuning $\Delta\Omega=0$ and under weak modulation depth (Fig.~\ref{fig:fig3}b). 
The oscillation frequency of the ZB dynamics is shown to be effectively tuned by the polarization rotation modulation and demonstrates clear $2\pi$ periodicity. 
The ZB oscillation frequency gets maximized at half of the $\Omega_\mathrm{R}$ when $\phi_y$ is near half-integers of $2\pi$, whereas the ZB phenomenon vanishes for $\phi_y$ being integers of $2\pi$. 

Next, we demonstrate the interplay between BO and ZB in Fig.~\ref{fig:fig3}c by turning on the modulation frequency detuning $\Delta\Omega\equiv\Omega-\Omega_\mathrm{R}$ while maintaining a weak modulation depth. 
The onset of $\Delta\Omega$ generates an Abelian electric field causing BO dynamics~\cite{yuan2016bloch}.
Therefore, under the coexisting Abelian and non-Abelian electric fields here, the measurement spectral dynamics exhibits the simultaneous presence of a strong BO line at $\Omega_{\mathrm{COM}}=\pm\Delta\Omega$ and the oscillatory ZB frequency trajectory. 
Moreover, both BO and ZB exhibit oscillation of their respective amplitudes that becomes most evident when the ZB frequency trajectory traverses the BO lines, indicating the strong interference between the two dynamics.

Finally, we enter the BO-dominant regime by further increasing the modulation depth (Fig.~\ref{fig:fig3}d).
While the BO frequency is solely determined by $\Delta\Omega$, its amplitude can be magnified by a stronger modulation depth.
Consequently, the ZB oscillations are strongly suppressed, but their amplitude interference with BO remains evident. 
Moreover, the residual ZB features asymmetric amplitudes in the upper and lower halves of the $\phi_y$ modulation cycle, 
which may stem from the asymmetric phase response of the modulated synthesizer serving as the HWP inside the ring.

The measured complicated two-dynamics interference becomes analytically intractable, but we can still obtain numerical validations beyond the first-order approximations. 
The simulations are shown next to the experimental data in Fig.~\ref{fig:fig3}b-d, where major features agree well with the measurement.  
Moreover, the simulations further reveal intriguing fine features, in particular the broadband interference fringes in Fig.~\ref{fig:fig3}c, which are absent in the data due to noise and decoherence. 
Such fringes again arise from the interplay between BO and ZB: the detuned modulation causes BO, while also replicating ZB oscillations centered at the BO frequencies. The interference among different copies of ZB leads to the fringes in Fig.~\ref{fig:fig3}c (details in SM Sec.~S6).

\vspace{0.2cm}

Based on a polarization multiplexed, time-modulated ring resonator, we experimentally synthesize non-Abelian electric fields and demonstrate their comprehensive control of spin-orbit coupling in the spin-resolved band structure measurements. Leveraging self-heterodyne detection, we observe the resulting dynamic phenomena of Zitterbewegung and its interference with Bloch oscillation under coexisting Abelian and non-Abelian electric fields. 
The realization here can be further extended to integrated platforms, in particular thin-film lithium niobate, where transverse-electric and transverse-magnetic polarization coupling can be dispersion engineered~\cite{hu2022mirror}.
By incorporating amplitude modulation, it is also possible to explore the non-Hermitian regime where non-Abelian fields can control complex energy winding and braiding~\cite{pang2024synthetic}.
In the presence of nonlinearity,  our setup could harbor soliton solutions to the mixed coupled nonlinear Schr\"{o}dinger equations, and may further support frequency-domain dynamic gauge fields for photonic emulation of quantum chromodynamics.

\vspace{0.4cm}

\emph{Acknowledgments}. 
We thank Yu Xia, Chao Xiang, Xudong Zhang, Xiaoqi Zhu for experimental support and Kun Ding, Jinbing Hu, and Zehai Pang for helpful discussions.

\vspace{0.4cm}

We acknowledge the support from the National Natural Science Foundation of China Excellent Young Scientists Fund (12222417), the Hong Kong Research Grants Council through General Research Fund (17300525), Early Career Scheme (27300924), Strategic Topics Grant (STG3/E-704/23-N), Collaborative Research Fund (C7015-24GF), Areas of Excellence Scheme (AoE/P-604/25-R), the Startup Fund of The University of Hong Kong, Ms.~Belinda Hung, the Asian Young Scientist Fellowship, the Croucher Foundation, the New Cornerstone Science Foundation through the Xplorer Prize.
\bibliographystyle{apsrev4-2}
\bibliography{reference}
\end{document}